\input epsf
\documentstyle[referee]{mn}
\begin{document}
\title[$IZ$ photometry of L dwarfs and the implications for
brown dwarf surveys]
{$IZ$ photometry of L dwarfs and the implications for
brown dwarf surveys}
\author[I.A. Steele \& L. Howells]
{I.A. Steele \& L. Howells \\
Astrophysics Research Institute, Liverpool John Moores University, 
Liverpool, CH41 1LD\\
}
\maketitle
\begin{abstract}
The $I-Z$ colour has been recently shown to be 
a good temperature indicator for
M dwarfs.  We present the first $IZ$ photometry of
a small sample of L dwarfs ranging in spectral type from L0.5V to
L6.0V.  We find that the $I-Z$ colour is not a good temperature
indicator for objects between L1V and L5V, such objects
having colours that overlap with mid M dwarfs.  
We attribute this
to the reduction in the strength of the TiO and VO bands in the L dwarfs
which are the dominant opacity source in the $I$ band for late M dwarfs. 
Beyond L5V,
$I-Z$ appears to be a reasonable indicator.   This has
important implications for the planning of optical surveys for cool
objects in clusters and the field.  For example $I-Z$ will
cease to be a good method of discriminating brown dwarfs in the Pleiades
below around $0.04 M_\odot$, and at around $0.075 M_\odot$ in the
Hyades and Praesepe.
\end{abstract}

\section{Introduction}

Kirkpatrick et al. (1999) have recently identified and classified the
first large sample of L dwarfs (objects cooler than M-dwarfs) from 
the 2-micron all sky survey (2MASS).  The optical spectra of these objects
are characterised by the disappearance of the TiO and VO bands
which dominate late M dwarfs, and their replacement by metallic
Hydrides and neutral alkali metals.  This would be expected to 
affect the optical colours one observes for such objects.  This is
important as many surveys for low mass objects (specifically brown dwarfs) 
in clusters have relied
on colours such as $V-I$ (e.g. Stauffer, Hamilton \& Probst 1994), 
$R-I$ (e.g. Jameson \& Skillen 1989, Hambly et al. 1999), 
and $I-Z$ (e.g  Pinfield et al. 1997, Cossburn et al. 1997, 
Zapatero Osorio et al. 1999).  The $I-Z$ 
colour has been particularly favoured of late, as it has been found to
be an excellent method of picking late cluster M dwarfs from the
field.  The lowest mass object so far detected in the Pleiades
using this technique is the $\sim$ L1V dwarf Roque 25 
which was recently identified by Martin et al. (1998).
It is therefore useful to see if the $I-Z$ colour
would remain useful as a method of picking out later $L$ dwarfs in cluster
fields.  For example at the distance of the Hyades 
an L3V dwarf would have $I\sim 19$ based on the pseudo-photometry 
(derived from flux calibrated spectra) presented by Kirkpatrick et al. (1999).
This is easily obtained with 2-m class telescopes, and therefore
an optical survey at $I$ and $Z$, at which wavelengths much larger
fields of view are generally available than the near infrared $JHK$ bands,
would appear to be an excellent way of finding such objects.  

In order to address this question of applicability to cluster surveys,
as well as the more general question of the use of $I-Z$ as a temperature 
indicator we have carried out $I_{\rm Harris}$ (hereafter $I_H$)
and $Z_{RGO}$ photometry of a small number of L dwarfs ranging in spectral type
from L0.5V to L6V.  This paper presents the results of that photometry,
and discusses the results in the context set out above.

\section{Observations}

Our observations were obtained using the 1.0-m Jacobus Kapteyn Telescope
(JKT), La Palma on the night of 1999 December 19.  The photometric conditions
were excellent, with relative humidity always below 5\% (giving 
good stability, especially in the $Z$ band) and no
cirrus or other cloud.  The lunar phase was near full, however this
does not affect the $I$ and $Z$ bands as much as the bluer optical bands.
Through the course of the night the seeing was reasonable, 
slowly varying between 1.1 and 1.3 arcsec FWHM.  This was sufficient
to adequately resolve the two of our program objects 
(2MASSW J0147334+345411 and 2MASSs J0850359+105716) that are
close ($\sim 2$ arcsec) binaries. 
Each object in the
programme was observed three times through each filter, with an
integration time of 600 seconds per observation (except for
the faintest object, 2MASSs J0850359+105716, where the integrations
were each 1200 seconds).  The standard field PG0918 (Landolt 1992) was
observed between each object (i.e. roughly once per 90 minutes) throughout
the night.

The CCD employed was a SITE 2048x2048 pixel array, and the
filters used were $I_H$ and $Z_{RGO}$.  These filters are
typically the ones used for the majority of $IZ$ surveys of
clusters so far carried out.
They 
have been calibrated using a TEK 1024x1024 CCD for a sample
of Landolt (1992) 
standards and M-dwarfs by Cossburn et al. (2000).  A comparison
of the measured 
quantum efficiency curves of the two CCDs shows that they are
identical to within $\sim 2$\% over the range 6000 - 10000 {\AA}.  The 
standards and M-dwarfs from Cossburn et al. (2000) are therefore
directly applicable to the newer SITE detector.
It is important to note that the Cossburn et al. (2000) calibration is
based on an assigned $I-Z$ colour of 0.0 for an unreddened A0 star and that
this is {\em different} to the Gunn $z$ and Sloan $z^\prime$ systems.

Data reduction was carried out in the usual manner, using a combination
of twilight fields and the median of the programme frames in the
appropriate bands to flat field and defringe the data. 

The results of our photometry are presented in Table 1.  Errors were
estimated from the dispersion of the measurements of each object.  
One object, 2MASSW J0918382+213406 has a very bright star roughly 1 arcminute
SW, making background estimation difficult due to scattered light.  This is
reflected in the greater photometric errors for this object compared to
the fainter 2MASSW J0913032+184150.

\begin{table*}
\caption{Photometry of 2MASS L-dwarfs.  Spectral types from Kirkpatrick 
et al. (1999).}
\begin{tabular}{lllll}
\hline
Object & Spectral Type &  $I_H$ & $Z_{RGO}$ & $I_H-Z_{RGO}$ \\
2MASSW J0147334+345311 & L0.5V  & $18.20  \pm 0.05$ & $17.40 \pm 0.05$ 
& $0.80 \pm 0.07$ \\
2MASSW J0918382+213406 & L2.5V  & $18.25  \pm 0.20$ & $17.75 \pm 0.10$ 
& $0.50 \pm 0.22$ \\
2MASSW J0913032+184150 & L3.0V  & $19.30  \pm 0.10$ & $18.60 \pm 0.05$ 
& $0.70 \pm 0.12$ \\
2MASSs J0850359+105716 & L6.0V  & $20.00  \pm 0.20$ & $18.70 \pm 0.20$ 
& $1.30 \pm 0.30$ \\
\hline
\end{tabular}
\end{table*}

\section{Discussion}

In Figure 1 we plot $I_{H}-Z_{RGO}$ (hereafter $I-Z$) versus spectral
type (Kirkpatrick et al. 1999) for our sample (circles).  Also
plotted as crosses are the M-dwarf data of Cossburn et al. (2000) and
their observation of DENIS-P J1228.2-1547, which Kirkpatrick et al. (1999)
classify as L5V.  Based purely on their 
M dwarf sample Cossburn et al. (2000) made the reasonable
claim that $I-Z$ was a good temperature indicator down
to $\sim 2000K$ (M9).  
However from the figure we see that between 
$\sim$ L1V and $\sim$ L5V the $I-Z$ colour of the L dwarfs 
is {\em bluer} than the late M-dwarfs, and in fact overlaps with the
mid-M dwarfs.  $I-Z$ is therefore {\em not}
a good temperature indicator in the range $\sim 0.5 - 1$.

\def\epsfsize#1#2{0.5#1}
\begin{figure}
\setlength{\unitlength}{1.0in}
\centering
\begin{picture}(3.0,3.0)(0,0)
\put(-0.4,0.0){\epsfbox[0 0 2 2]{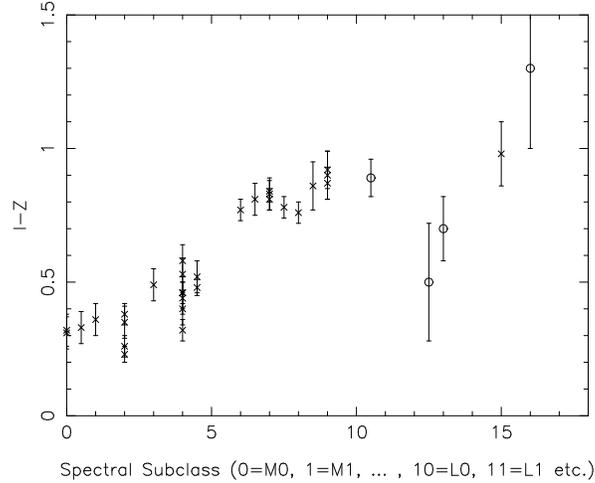}}
\end{picture}
\caption{Spectral Class, $I_H-Z_{RGO}$ diagram for our observations of
L-dwarfs (circles).  Also plotted (crosses) 
are the M-dwarf and DENIS-P J1228.2-1547
data of Cossburn et al. (2000)}
\end{figure}

The reason for the relatively blue $I-Z$ colour of the early-mid L dwarfs
can be understood by examining the spectra of Kirkpatrick et al. (1999).
A comparison of the L0V spectra with (for example) 
the L3V spectra shows that the
TiO and VO band strengths are much reduced in the L3V objects.  These 
opacity sources are especially strong between 7800-8000 {\AA} (VO) and
8400-8600 {\AA} (TiO), i.e. in the region where the $I$ band
filter transmission is greatest.  Overlaying the spectra
normalized at the pseudo-continuum point at $\sim 8250$ {\AA}
shows that the regions longward of $\sim 8600$ {\AA} (where the
$Z$ band transmission is greatest) overlap 
well, but a significant flux excess for the L3V object shorter than
this wavelength.  Therefore such objects will appear bluer than earlier
objects in $I-Z$.  For objects later than $\sim$ L5V $I-Z$ again
becomes a reasonable temperature indicator.  Examination of the
spectra shows that this is simply due to the extremely cool
temperature giving a very steep spectral slope through $I$ and $Z$ 
which `overwhelms' the effect of the lack of opacity in the
$I$ band.

As stated previously, $I-Z$ has recently become the favoured
colour for cluster brown dwarf searches, often with considerable
success (e.g. Cossburn  et al. 1997, Zapatero Osorio et al. 1999).  However
from Figure 1 it appears that if we wish to find cluster L dwarfs
(which compared with field objects 
have the advantage of known distance, age and metallicity,
making mass derivations via. a comparison with isochrones feasible),
the $I-Z$ colour would not be appropriate.  To confirm this in Figure 2
we plot the combined $I,I-Z$ diagram for the four fields containing
our L dwarfs.  It is apparent that only the earliest and
latest L dwarf would be identified from this diagram as
a potentially interesting object, the middle two objects
overlapping with the background objects (which may be
distant M dwarfs, even more distant M giants or extragalactic)
as would be expected from their bluer colours.  It is therefore
apparent that $I,Z$ searches for L dwarfs in clusters (and the field) 
will only be sensitive to objects earlier than $\sim$L1V and later than 
$\sim$L5V.
Using the objects listed in
Kirkpatrick et al. (1999) which have measured parallaxes
to define an absolute $I$ magnitude, spectral type relation 
for the L dwarfs, this indicates that
$I-Z$ will not be a useful colour in the absolute magnitude range
$M_I \sim 15 - 17$.  Assuming a Pleiades age of $\sim 120 $Myr 
and using an extension (Baraffe, priv. comm.) of the Lyon Group 
absolute magnitude-mass relationship presented by 
Baraffe et al. (1998) this
implies that for that cluster objects with masses lower than
$\sim 0.04M_\odot$ will be difficult to pick out with $IZ$ photometry.
This is consistent with the lowest mass (spectroscopically
confirmed) object so far found using that
technique (Roque 25) which has a spectral type
of $\sim$ L1V ($M\sim 0.04M_\odot$) 
(Martin et al. 
1998).
For older clusters such as the Hyades and Praesepe (age $\sim 1$ Gyr), 
objects in the range $\sim 0.075 - 0.06 M_{\odot}$ will be difficult to
detect using the $I-Z$ colour.  This is consistent with
the lowest mass objects detected in Praesepe with this
technique having $I \sim 21.5$ (Pinfield et al. 1997, Magazzu et al. 1998),
corresponding to $M_I \sim 15$.

\def\epsfsize#1#2{0.49#1}
\begin{figure}
\setlength{\unitlength}{1.0in}
\centering
\begin{picture}(3.0,4.6)(0,0)
\put(-0.4,0.0){\epsfbox[0 0 2 2]{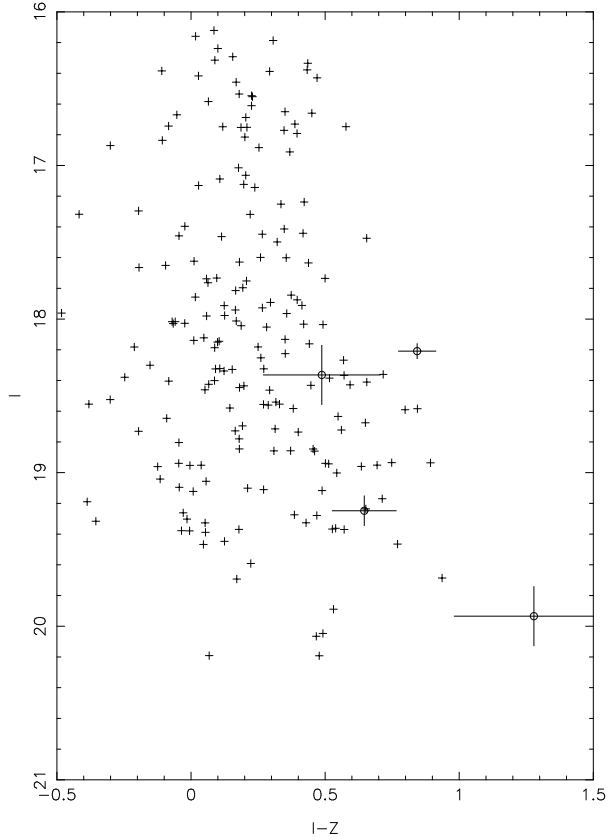}}
\end{picture}
\caption{$I,I-Z$ diagram for all of the objects in our observed fields.
The L-dwarfs are indicated by circles with error bars 
and the other objects by crosses.  Note that the photometry is based
on a single observation of each field, and therefore differs
slightly from the average value presented in Table 1.}
\end{figure}

\section{Conclusions}

We have presented observations that show that the $I-Z$ 
colour is not a good temperature indicator for
L dwarfs between L1V and L5V, these objects having $I-Z$ colours
which overlap with mid-late M dwarfs.  We attribute this to
the decreasing blanketing of the $I$ band flux in L dwarfs due
to the absence of strong TiO and VO bands in their spectra.  This
imposes limits on the use of $I-Z$ as an indicator of very cool
objects in cluster brown dwarf searches at around
$0.04 M_\odot$ for the Pleiades and $0.075 M_\odot$ for
Praesepe and the Hyades.  For objects of lower mass than this
near infrared ($JHK$) surveys should be considered.
  
\section*{Acknowledgements}

Data reduction for this paper was carried out on the
Liverpool John Moores STARLINK node. 
The JKT is operated by
the ING on behalf of the UK Particle Physics and
Astronomy Research Council (PPARC) at the ORM Observatory, La Palma.   
We are pleased to thank Rachel Curran and the technical staff
of the ING
for their assistance at the telescope.  We gladly 
acknowledge the postscript genius of Andrew Newsam
for the addition of error bars to Fig. 2.
LH acknowledges a PPARC research studentship.

\end{document}